\documentclass[prl,aps,nofootinbib,twocolumn]{revtex4}
\usepackage{graphicx,aas_macros}

\newcommand{\f}{\frac}
\newcommand{\Msun}{\,\rm M_\odot}
\newcommand{\mDM}{m_\chi}
\newcommand{\sigv}{\langle \sigma v \rangle}
\newcommand{\Vmax}{V_{\rm max}}
\newcommand{\rVmax}{r_{\rm Vmax}}
\newcommand{\Vpeak}{V_{\rm peak}}
\newcommand{\rVpeak}{r_{\rm Vpeak}}
\newcommand{\Fermi}{\textit{Fermi}}

\newcommand{\val}[3]{$#1\;(^{#2}_{#3})$}
\newcommand{\vale}[3]{$#1^{#2}_{#3}$}

\newcommand{\lum}{\mathcal{L}}
\newcommand{\F}{\mathcal{F}}
\newcommand{\Ftot}{\mathcal{F}_{\rm tot}}
\newcommand{\Fc}{\mathcal{F}_c}

\renewcommand{\tabcolsep}{0.15in}

\begin{document}

\title{The Dark Matter Annihilation Signal from Dwarf Galaxies and Subhalos}
\author{Michael Kuhlen}
\affiliation{Institute for Advanced Study, School of Natural Science\\
  Einstein Lane, Princeton, NJ 08540 \\
  Email address: mqk@ias.edu}

\begin{abstract}
Dark Matter annihilation holds great potential for directly probing
the clumpiness of the Galactic halo that is one of the key predictions
of the Cold Dark Matter paradigm of hierarchical structure
formation. Here we review the $\gamma$-ray signal arising from dark
matter annihilation in the centers of Galactic subhalos. We consider
both known Galactic dwarf satellite galaxies and dark clumps without a
stellar component as potential sources. Utilizing the \textit{Via
  Lactea II} numerical simulation, we estimate fluxes for 18 Galactic
dwarf spheroidals with published central densities. The most promising
source is Segue 1, followed by Ursa Major II, Ursa Minor, Draco, and
Carina. We show that if any of the known Galactic satellites can be
detected, then at least ten times more subhalos should be visible,
with a significant fraction of them being dark clumps.
\end{abstract}

\maketitle

\section{Introduction}

A decade has gone by since the emergence of the ``Missing Satellite
Problem'' \cite{Klypin1999, Moore1999}, which refers to the apparent
discrepancy between the observed number of Milky Way satellite
galaxies, 23 by latest count
\cite{Mateo1998,Willman2005a,Willman2005b,Belokurov2006,Zucker2006a,Zucker2006b,Sakamoto2006,Belokurov2007,Irwin2007,Walsh2007},
and the predicted number of dark matter (DM) subhalos that should be
orbiting in the Milky Way's halo. The latest cosmological numerical
simulations \cite{Diemand2008, Springel2008, Stadel2008} resolve close
to 100,000 individual self-bound clumps of DM within the Galactic
virial volume -- remnants of the hierarchical build-up of the Milky
Way's DM halo. A consensus seems to be emerging that this discrepancy
is not a short-coming of the otherwise tremendously successful Cold
Dark Matter (CDM) hypothesis \cite{White1978,Blumenthal1984}, but
instead reflects the complicated baryonic physics that determines
which subhalos are able to host a luminous stellar component and which
aren't
\cite{Dekel1986,Bullock2000,Kravtsov2004,Mayer2006,Madau2008,Koposov2009,Maccio2009}. If
this explanation is correct, then an immediate consequence is that the
Milky Way dark halo should be filled with clumps on all scales down to
the CDM free-streaming scale of $10^{-12}$ to $10^{-4} \Msun$
\cite{Profumo2006,Bringmann2009}. At the moment there is little
empirical evidence for or against this prediction, and this has
motivated searches for new signals that could provide tests of this
hypothesis, and ultimately help to constrain the nature of the DM
particle.

One of the most promising such signals is DM annihilation
\cite{Bergstroem1999}. In regions of sufficiently high density, for
example in the centers of Galactic subhalos, the DM pair annihilation
rate might become large enough to allow for a detection of neutrinos,
energetic electrons and positrons, or $\gamma$-ray photons, which are
the by-products of the annihilation process. This is one of the few
ways in which the dark sector can be coupled to ordinary matter and
radiation amenable to astronomical observation. Belying its commonly
used name of ``indirect detection'', DM annihilation is really the
only way we can hope to directly probe the clumpiness of the Galactic
DM distribution. One could argue that it is a more ``direct'' method
than trying to constrain DM clumpiness from its effects on strong
gravitational lensing (see Zackrisson \& Riehm's contribution in this
special edition), or from the kinematics of stars orbiting in
DM-dominated potentials \cite{Strigari2008}, or from perturbations of
cold stellar structures like globular cluster tidal streams
\cite{Ibata2002,Johnston2002,Penarrubia2006,Siegal-Gaskins2008b} or
the heating of the Milky Way's stellar disk
\cite{Toth1992,Read2008,Kazantzidis2009}, although all of these are
also worthwhile approaches to take.

The only trouble with the DM annihilation signal is that so far there
have been no undisputed claims of its detection. Recently there have
been several reports of ``anomalous'' features in the local cosmic ray
flux: the PAMELA satellite reported an increasing positron fraction at
energies between 10 and 100 GeV \cite{Adriani2009}, where standard
models of cosmic ray propagation predict a decreasing fraction; the
ATIC \cite{Chang2008} and PPB-BETS \cite{Torii2008} balloon-borne
experiments reported a surprisingly large total electron and positron
flux at $\sim 500$ GeV, although recent \Fermi\ \cite{Abdo2009} and
H.E.S.S. data \cite{Aharonian2009} appear to be inconsistent with
it. Either of these cosmic ray anomalies might be the long sought
after signature of local DM annihilation. However, since the currently
available data can equally well be explained by conventional
astrophysical sources (e.g. nearby pulsars or supernova remnants),
they hardly provide incontrovertible evidence for DM annihilation. The
next few years hold great potential for progress, since the recently
launched \textit{Fermi Gamma-ray Space Telescope} will conduct a blind
survey of the $\gamma$-ray sky at unprecedented sensitivity, energy
extent, and angular resolution. At the same time, Atmospheric Cerenkov
Telescopes, such as H.E.S.S., VERITAS, MAGIC, and STACEE, are greatly
increasing their sensitivity, and have only recently begun to search
for a DM annihilation signal from the centers of nearby dwarf
satellite galaxies
\cite{Aharonian2008,Hui2008,Albert2008,Aliu2009,Driscoll2008}.

The purpose of this paper is to provide an overview of the potential
DM annihilation signal from individual Galactic DM subhalos, either as
dwarf satellite galaxies or as dark clumps. It does not cover a number
of very interesting and closely related topics, which are actively
being researched and deserve to be examined in equal detail. These
include the diffuse flux from Galactic substructure and its
anisotropies \cite[e.g.][]{Siegal-Gaskins2008, Ando2009, Fornasa2009},
the relative strength of the signal from individual subhalos compared
with that from the Galactic Center or an annulus around it
\cite{Stoehr2003, Springel2008b}, the effect of a nearby DM subhalo on
the amplitude and spectrum of the local high energy electron and
positron flux \cite{Brun2009,Kuhlen2009}, and the role of the
Sommerfeld enhancement \cite{Arkani-Hamed2009} on the DM annihilation
rate and its implications for substructure signals
\cite{Lattanzi2009,Pieri2009b,Kuhlen2009b}.

This paper is organized as follows: we first review the basic physics
of DM annihilation, briefly touching on the relic density calculation,
the ``WIMP miracle'', DM particle candidates, and, in more detail, the
sources of $\gamma$-rays from DM annihilation. In the following
section we review what numerical simulations have revealed about the
basic properties of DM subhalos that are relevant for the annihilation
signal. We go on to consider known Milky Way dwarf spheroidal galaxies
as sources, using the \textit{Via Lactea II} simulation to infer the
most likely annihilation fluxes from published values of the dwarfs'
central masses. Next we discuss the possibility of a DM annihilation
signal from dark clumps, halos that have too low a mass to host a
luminous stellar component. Lastly, we briefly discuss the role of the
substructure boost factor for the detectability of individual DM
subhalos.

\section{Dark Matter Annihilation}

If DM is made up of a so-called ``thermal relic'' particle\footnote{An
  alternative DM candidate is the axion, a non-thermal relic particle
  motivated as a solution to the strong CP problem \cite{Turner1990}.
  Since it doesn't produce an annihilation signal today, we don't
  further consider it here.}, its abundance today is set by its
annihilation cross section in the early universe. The thermal relic
abundance calculation relating today's abundance of DM to the
properties of the DM particle (its mass and annihilation cross
section) is straightforward and elegant, and has been described in
pedagogical detail previously \cite{Kolb1990,Jungman1996,Bertone2005}.
We briefly summarize the story here.

At sufficiently early times, the DM particles are in thermal
equilibrium with the rest of the universe. As long as they remain
relativistic ($T \gg \mDM$), their creation and destruction rates are
balanced, and hence their co-moving abundance remains constant. Once
the universe cools below the DM particle's rest-mass ($T < \mDM$), its
equilibrium abundance is suppressed by a Boltzmann factor
$\exp(-\mDM/T)$. If equilibrium had been maintained until today, the
DM particles would have completely annihilated away. Instead the
expansion of the universe comes to the rescue and causes the DM
particles to fall out of equilibrium once the expansion rate (given by
$H(a)$, the Hubble constant at cosmological scale factor $a$) exceeds
the annihilation rate $\Gamma(a)=n\sigv$, i.e. when DM particles can
no longer find each other to annihilate. The co-moving number density
of DM particles is then fixed at a ``freeze-out'' temperature that
turns out to be approximately $T_f \simeq \mDM/20$, with only a weak
additional logarithmic dependence on the mass and cross section of the
DM particle. A back of the envelope calculation results in the
following relation between $\Omega_\chi$, the relic mass density in
units of the critical density of the universe $\rho_{\rm crit} =
3H_0^2/8\pi G$, and $\sigv$, the thermally averaged velocity-weighted
annihilation cross section:
\begin{equation}
\omega_\chi = \Omega_\chi h^2 = \f{3 \times 10^{-27} \; {\rm cm}^3 \; {\rm s}^{-1}}{\sigv}.
\end{equation}
Note that this relation is independent of $\mDM$. The WMAP satellite's
measurement of the DM density is $\omega_\chi = 0.1131 \pm 0.0034$
\cite{Hinshaw2009}, implying a value of
\begin{equation}
\sigv \approx 3 \times 10^{-26} \; {\rm cm}^3 \; {\rm s}^{-1}.
\end{equation}
A more accurate determination of $\sigv$ must rely on a numerical
solution of the Boltzmann equation in an expanding universe, taking
into account the full temperature dependence of the annihilation rate,
including the possibilities of resonances and co-annihilations into
other, nearly degenerate dark sector particles
\cite[e.g.][]{Griest1991,Gondolo1991}. It is a remarkable coincidence
that this value of $\sigv$ is close to what one expects for a cross
section set by the weak interaction. This is the so-called ``WIMP
miracle'', and it is the main motivation for considering weakly
interacting massive particles (WIMPs) as prime DM candidates.

The Standard Model of particle physics actually provides one class of
WIMPs, massive neutrinos. Although neutrinos thus constitute a form of
DM, they cannot make up the bulk of it, since their small mass, $\sum
m_\nu < 0.63$ eV \cite{Hinshaw2009}, implies a cosmological mass
density of only $\omega_\nu = 7.1 \times 10^{-3}$. The attention thus
turns to extensions of the Standard Model, which themselves are
theoretically motivated by the hierarchy problem (the enormous
disparity between the weak and Planck scales) and the quest for a
unification of gravity and quantum mechanics. The most widely studied
class of such models consists of supersymmetric extensions of the
Standard Model, although models with extra dimensions have received a
lot of attention in recent years as well. Both of these approaches
offer good DM particle candidates: the lightest supersymmetric
particle (LSP), typically a \textit{neutralino} in R-parity conserving
supersymmetry, and the lightest Kaluza-Klein particle (LKP), typically
the $B^{(1)}$ particle, the first Kaluza-Klein excitation of the
hypercharge gauge boson, in Universal Extra Dimension models. For much
more information, we recommend the comprehensive recent review of
particle DM candidates by Bertone, Hooper \& Silk \cite{Bertone2005}.

\begin{figure}
\includegraphics[width=\columnwidth]{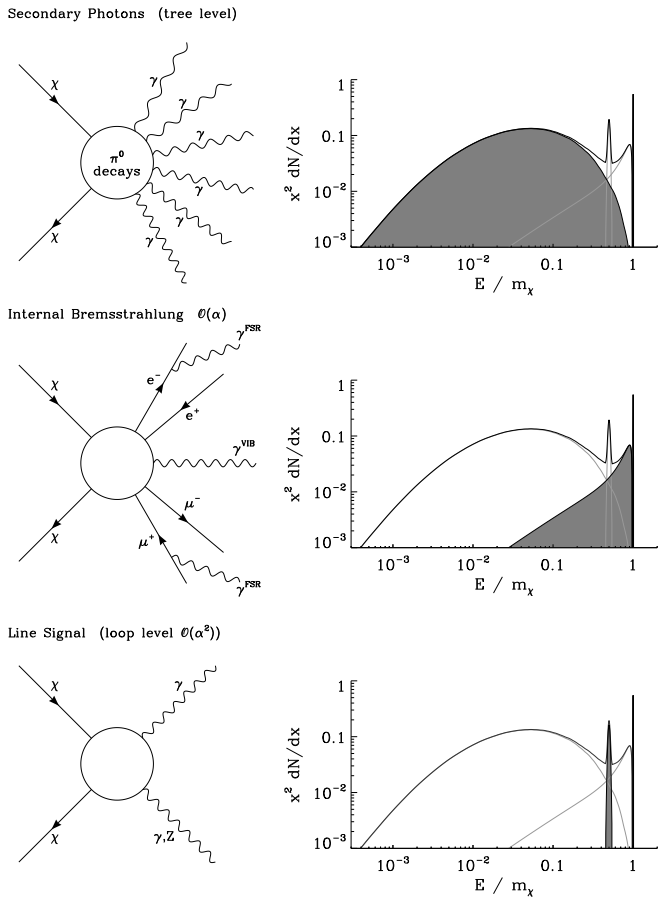}
\caption{A schematic of the different sources and energy distributions
  of $\gamma$-rays from WIMP annihilation. \textit{Top:} Secondary
  photons arising from the decay of neutral pions produced in the
  hadronization of primary annihilation products. \textit{Middle:}
  Internal bremsstrahlung photons associated with charged annihilation
  products, either in the form of final state radiation (FSR) from
  external legs or as virtual internal bremsstrahlung (VIB) from the
  exchange of virtual charged particles. \textit{Bottom:}
  Mono-chromatic line signals from the prompt annihilation into two
  photons or a photon and $Z$ boson. This process occurs only at loop
  level, and hence is typically strongly suppressed.}
\label{fig:gamma}
\end{figure}

The direct products of the annihilation of two DM particles are
strongly model dependent. Typical channels include annihilations into
charged leptons ($e^+e^-, \mu^+\mu^-, \tau^+\tau^-$), quark-antiquark
pairs, and gauge and Higgs bosons ($W^+W^-, Z, h$). In the end,
however, the decay and hadronization of these annihilation products
results in only three types of emissions: (i) high energy neutrinos,
(ii) relativistic electrons and protons and their anti-particles, and
(iii) $\gamma$-ray photons. Additional lower energy photons can result
from the interaction of the relativistic electrons with magnetic
fields (synchrotron radiation), with interstellar material
(bremsstrahlung), and with the CMB and stellar radiation fields
(inverse Compton scattering). In the following we will focus on the
$\gamma$-rays, since they are likely the strongest signal from
Galactic DM substructure. $\gamma$-rays are produced in DM
annihilations in three ways (see accompanying Fig.~\ref{fig:gamma})
\begin{itemize}
\item[(i)] Since the DM particle is neutral, there is no direct
  coupling to photons. Nevertheless, copious amounts of secondary
  $\gamma$-ray photons can be produced through the decay of neutral
  pions, $\pi^0 \rightarrow \gamma \gamma$, arising in the
  hadronization of the primary annihilation products. Since the DM
  particles are non-relativistic, their annihilation results in a pair
  of mono-energetic particles with energy equal to $\mDM$, which
  fragment and decay into $\pi$-meson dominated ``jets''. In this way
  a single DM annihilation event can produce several tens of
  $\gamma$-ray photons. The result is a broad spectrum with a cutoff
  around $\mDM$.

\item[(ii)] An important additional contribution at high energies ($E
  \lesssim \mDM$) arises from the internal bremsstrahlung process
  \cite{Bringmann2008}, which may occur with any charged annihilation
  product. One can distinguish between final state radiation, in which
  the photon is radiated from an external leg, and virtual internal
  bremsstrahlung, arising from the exchange of a charged virtual
  particle. Note that neither of these processes requires an external
  electromagnetic field (hence the name \textit{internal}
  bremsstrahlung). The resulting $\gamma$-ray spectrum is peaked
  towards $E \sim \mDM$ and exhibits a sharp cutoff. Although it is
  suppressed by one factor of the coupling $\alpha$ compared to pion
  decays, it can produce a distinctive spectral feature at high
  energies. This could aide the confirmation of a DM annihilation
  nature of any source and might allow a direct determination of
  $\mDM$.

\item[(iii)] Lastly, it is possible for DM particles to directly
  produce $\gamma$-ray photons, but one has to go to loop-level to
  find contributing Feynman diagrams, and hence this flux is typically
  strongly suppressed by two factors of $\alpha$ (although exceptions
  exist \cite{Gustafsson2007}). On the other hand, the resulting
  photons would be mono-chromatic, and a detection of such a line
  signal would provide strong evidence of a DM annihilation origin of
  any signal. While annihilations directly into two photons, $\chi\chi
  \rightarrow \gamma\gamma$, would produce a narrow line at $E =
  \mDM$, in some models it is also possible to annihilate into a
  photon and a $Z$ boson, $\chi\chi \rightarrow \gamma Z$, and this
  process would result in a somewhat broadened (due to the mass of the
  $Z$) line at $E \sim \mDM (1 - m_Z^2/4\mDM^2$).

\end{itemize}

The relative importance of these three $\gamma$-ray production
channels and the resulting spectrum $dN_\gamma/dE$ depend on the
details of the DM particle model under consideration. For any given
model, realistic $\gamma$-ray spectra can be calculated using
sophisticated and publicly available computer programs, such as the
\texttt{PYTHIA} Monte-Carlo event generator \cite{Sjoestrand1994},
which is also contained in the popular \texttt{DarkSUSY} package
\cite{Gondolo2004}.

\section{Dark Matter Substructure as Discrete $\gamma$-ray Sources}

DM subhalos as individual discrete $\gamma$-ray sources hold great
potential for providing a ``smoking gun'' signature of DM annihilation
\cite{Bergstroem1999,Calcaneo-Roldan2000,Baltz2000,Tasitsiomi2002,Stoehr2003,Aloisio2004,Evans2004,Koushiappas2004,Koushiappas2006,Diemand2007a,Pieri2008,Kuhlen2008,Strigari2008b}.
Compared to diffuse $\gamma$-ray annihilation signals, these discrete
sources should be easier to distinguish from astrophysical backgrounds
and foregrounds \cite{Baltz2007}, since a) typical astrophysical
sources of high energy $\gamma$-rays, such as pulsars and supernova
remnants, are very rare in dwarf galaxies, owing to their
predominantly old stellar populations, b) the DM annihilation flux
should be time-independent, c) angularly extended, and d) not exhibit
any (or only very weak) low energy emission due to the absence of
strong magnetic fields or stellar radiation fields.

We can distinguish between DM subhalos hosting a Milky Way dwarf
satellite galaxy and dark clumps that, for whatever reason, don't host
a luminous stellar population, or one that is too faint to have been
detected up to now. Before we go on to discuss the prospects of
detecting a DM annihilation signal from these two classes of sources,
we review the basic properties of DM subhalos common to both.

% DM density profile, total luminosity, luminosity profile

\begin{figure}
\includegraphics[width=\columnwidth]{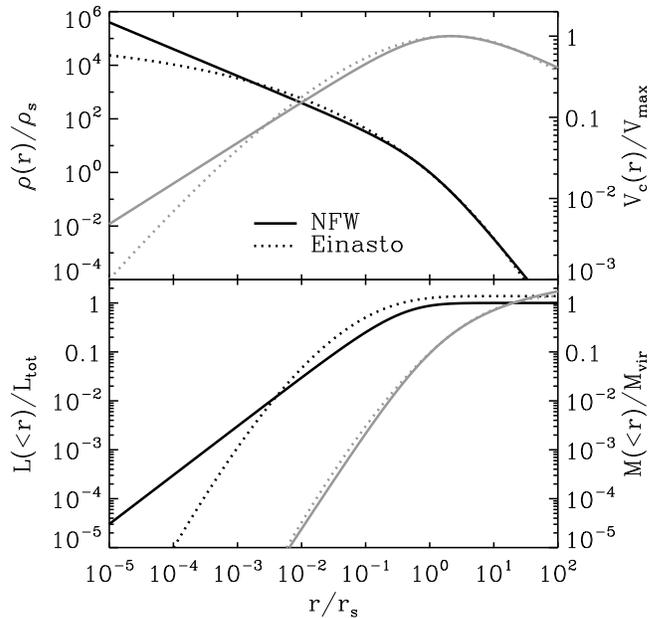}
\caption{A comparison of NFW and Einasto ($\alpha=0.17$) radial
  profiles of density (top, dark lines, left axis), circular velocity
  (top, light lines, right axis), enclosed annihilation luminosity
  (bottom, dark lines, left axis), enclosed mass (bottom, light lines,
  right axis). The density profiles have been normalized to have the
  same $\Vmax$ and $\rVmax$.}
\label{fig:profiles}
\end{figure}

Numerical simulations have shown that pure DM (sub)halos have density
profiles that are well described by a Navarro, Frenk \& White (NFW)
\cite{Navarro1997} profile over a wide range of masses
\cite{Maccio2007,Diemand2004},
\begin{equation}
\rho_{\rm NFW}(r) = \f{4 \rho_s}{(r/r_s) (1+r/r_s)^2}.
\end{equation}
The parameter $r_s$ indicates the radius at which the logarithmic
slope $\gamma(r) \equiv \f{d\ln\rho}{d\ln r} = -2$, and $\rho(r_s) =
\rho_s$. The very highest resolution simulations have recently
provided some indications of a flattening of the density profile in
the innermost regions \cite{Navarro2008,Stadel2008}. In this case a
so-called Einasto profile may provide a better overall fit,
\begin{equation}
\rho_{\rm Einasto}(r) = \rho_s \exp{\left[-\f{2}{\alpha}\left( \left(\f{r}{r_s}\right)^\alpha - 1 \right)\right]}.
\end{equation}
Here the additional parameter $\alpha$ governs how fast the profile
rolls over, and has been found to have a value of $\alpha \approx 0.17
\pm 0.03$ in numerical simulations \cite{Navarro2008}. Note that the
two density profiles actually do not differ very much until $r \ll
r_s$ (cf. top panel of Figure~\ref{fig:profiles}). Simulated DM halos
are of course not perfectly spherically symmetric, and instead
typically exhibit prolate or triaxial iso-density contours that become
more elongated towards the center \cite{Allgood2006}. The degree of
prolateness decreases with mass, and galactic subhalos have axis
ratios of $\gtrsim 0.7$ \cite{Kuhlen2007}.

The ``virial'' radius $R_{\rm vir}$ of a halo is defined as the radius
enclosing a mean density equal to $\Delta_{\rm vir} \rho_0$, where
$\Delta_{\rm vir} \approx 389$ \cite{Bryan1998} and $\rho_0$ is the
mean density of the universe. The corresponding virial mass $M_{\rm
  vir}$ is the mass within $R_{\rm vir}$, and a halo's concentration
can then be defined as $c=R_{\rm vir}/r_s$. While these quantities are
well defined for isolated halos and commonly used in analytic models,
they are somewhat less applicable to galactic subhalos, since the
outer radius of a subhalo is set by tidal truncation, which depends on
the subhalo's location within its host halo. Furthermore, in numerical
simulations it is difficult to resolve $r_s$ in low mass subhalos. For
this reason we prefer to work with $\Vmax$, the maximum of the
circular velocity curve $V_c(r)^2 = G M(<r)/r$ and a proxy for a
subhalo's mass, and $\rVmax$, the radius at which $\Vmax$ occurs.
These quantities are much more robustly determined for subhalos in
numerical simulations than $(M,c)$. Note that even $(\Vmax, \rVmax)$
can be affected by tidal interactions with the host halo, especially
for subhalos close to the host halo center. For this reason we also
sometimes consider $\Vpeak$, the largest value of $\Vmax$ that a
subhalo ever acquired during its lifetime (i.e. before tidal stripping
began to lower its $\Vmax$) and $\rVpeak$, the corresponding radius.

Since DM annihilation is a two body process, its rate is proportional
to the square of the local density, and the annihilation
``luminosity'' is given by the volume integral of $\rho(r)^2$,
\begin{equation}
\lum(<r) \equiv \int_0^r \rho^2 \; dV.
\end{equation}
$\lum$ has dimensions of (mass)$^2$ (length)$^{-3}$, and we express it
in units of $\Msun^2$ pc$^{-3}$. In order to convert to a conventional
luminosity, one must multiply by a particle physics term,
\begin{equation}
L = c^2 \f{\sigv}{\mDM} \lum,
\end{equation}
where $\mDM$ is the mass of the DM particle and $\sigv$ the thermally
averaged velocity-weighted annihilation cross section discussed in the
previous section. This is the \textit{total} luminosity, but we are
interested here only in the fraction emitted as
$\gamma$-rays. Furthermore, a given detector is only sensitive to
$\gamma$-rays above a threshold energy of $E_{\rm th}$ and below a
maximum energy of $E_{\rm max}$. In that case the effective
$\gamma$-ray luminosity is
\begin{equation}
L^{\rm eff}_\gamma = \left[ \f{\sigv}{2\mDM^2} \int_{E_{\rm th}}^{E_{\rm max}}
  \!\!\! E \f{dN_\gamma}{dE} dE \right] \lum,
\label{eq:Lconversion}
\end{equation}
where $dN_\gamma/dE$ is the spectrum of $\gamma$-ray photons produced
in a single annihilation event.

A comparison of the enclosed luminosity and mass profiles is shown in
the bottom panel of Figure~\ref{fig:profiles}. Clearly, $\lum$ is much
more centrally concentrated than $M$: $\sim 90\%$ of the total
luminosity is produced within $r_s$, compared with only 10\% of the
total mass. In terms of $(\Vmax, \rVmax)$, the total luminosity of a
halo is given by
\begin{equation}
\lum = f \f{\Vmax^4}{G^2 \rVmax},
\end{equation}
where $f$ is an $\mathcal{O}(1)$ numerical factor that depends on the
shape of the density profile; for an NFW profile $f=1.227$, and for an
$\alpha=0.17$ Einasto profile $f=1.735$. In physical units, the total
annihilation luminosity is
\begin{equation}
\lum = \begin{array}{c}1.1 \\ 1.5\end{array} \times 10^7 \; \Msun^2 \; {\rm pc}^{-3} \left( \f{\Vmax}{20 \,{\rm km} \; {\rm s}^{-1}} \right)^4 \left( \f{\rVmax}{1 {\rm kpc}} \right)^{-1},
\label{eq:Lhalo}
\end{equation}
for NFW (top) and $\alpha=0.17$ Einasto (bottom). Note that even
though the slope of the Einasto profile is shallower than NFW in the
very center, the total luminosity exceeds that of an NFW halo with the
same $(\Vmax, \rVmax)$. This is due to the fact that the Einasto
profile rolls over less rapidly than the NFW profile, and actually has
slightly higher density than NFW between $r_s$ and a cross-over point
at $\sim 10^{-3} r_s$.

\section{Milky Way Dwarf Spheroidal Galaxies}

\renewcommand{\arraystretch}{1.5}

\begin{table*}
\begin{tabular}{lcccccc}
Name & $D$ & $M_{0.3}$ & $\Vmax$ & $\rVmax$ & $\Vpeak$ & $\rVpeak$ \\
     & [kpc] & [$10^7 M_\odot$] & [km\,s$^{-1}$] & [kpc] & [km\,s$^{-1}$] & [kpc] \\
\hline \hline
Segue 1 & 23 & \vale{1.58}{+3.30}{-1.11} & \val{ 10}{ 17}{8.4} & \val{0.43}{0.89}{0.29} & \val{ 26}{ 55}{ 13} & \val{2.4}{ 33}{1.4}\\
Ursa Major II & 32 & \vale{1.09}{+0.89}{-0.44} & \val{ 13}{ 17}{ 11} & \val{0.59}{0.89}{0.31} & \val{ 27}{ 33}{ 17} & \val{3.3}{ 14}{2.4}\\
Wilman 1 & 38 & \vale{0.77}{+0.89}{-0.42} & \val{8.3}{ 11}{7.5} & \val{0.38}{0.62}{0.29} & \val{ 15}{ 27}{ 10} & \val{2.0}{3.9}{0.90}\\
Coma Berenices & 44 & \vale{0.72}{+0.36}{-0.28} & \val{9.1}{ 12}{8.2} & \val{0.42}{0.62}{0.31} & \val{ 15}{ 25}{ 11} & \val{1.9}{3.4}{0.97}\\
Ursa Minor & 66 & \vale{1.79}{+0.37}{-0.59} & \val{ 18}{ 21}{ 15} & \val{0.81}{1.8}{0.61} & \val{ 30}{ 56}{ 21} & \val{3.8}{9.7}{2.8}\\
Draco & 80 & \vale{1.87}{+0.20}{-0.29} & \val{ 19}{ 22}{ 17} & \val{0.86}{2.4}{0.81} & \val{ 28}{ 37}{ 26} & \val{3.8}{ 32}{2.4}\\
Sculptor & 80 & \vale{1.20}{+0.11}{-0.37} & \val{ 13}{ 15}{ 12} & \val{0.64}{1.0}{0.54} & \val{ 20}{ 25}{ 16} & \val{2.9}{5.6}{1.6}\\
Sextans & 86 & \vale{0.57}{+0.45}{-0.14} & \val{9.7}{ 12}{8.5} & \val{0.52}{0.89}{0.37} & \val{ 14}{ 19}{ 11} & \val{1.6}{3.0}{0.97}\\
Carina & 101 & \vale{1.57}{+0.19}{-0.10} & \val{ 17}{ 22}{ 16} & \val{1.00}{2.3}{0.69} & \val{ 30}{ 42}{ 24} & \val{3.8}{ 32}{3.3}\\
Ursa Major I & 106 & \vale{1.10}{+0.70}{-0.29} & \val{ 14}{ 17}{ 13} & \val{0.84}{1.3}{0.61} & \val{ 20}{ 30}{ 16} & \val{3.2}{6.8}{1.6}\\
Fornax & 138 & \vale{1.14}{+0.09}{-0.12} & \val{ 15}{ 16}{ 14} & \val{1.1}{1.3}{0.64} & \val{ 20}{ 24}{ 18} & \val{3.0}{6.1}{1.9}\\
Hercules & 138 & \vale{0.72}{+0.51}{-0.21} & \val{ 11}{ 14}{9.4} & \val{0.69}{1.1}{0.45} & \val{ 14}{ 20}{ 12} & \val{1.9}{3.8}{1.2}\\
Canes Venatici II & 151 & \vale{0.70}{+0.53}{-0.25} & \val{ 11}{ 13}{8.9} & \val{0.67}{1.1}{0.44} & \val{ 14}{ 19}{ 11} & \val{1.8}{3.7}{1.1}\\
Leo IV & 158 & \vale{0.39}{+0.50}{-0.29} & \val{5.0}{7.2}{4.2} & \val{0.35}{0.57}{0.22} & \val{6.7}{ 10}{5.0} & \val{0.84}{1.7}{0.48}\\
Leo II & 205 & \vale{1.43}{+0.23}{-0.15} & \val{ 18}{ 21}{ 16} & \val{1.5}{2.1}{0.93} & \val{ 24}{ 28}{ 19} & \val{4.1}{8.2}{2.4}\\
Canes Venatici I & 224 & \vale{1.40}{+0.18}{-0.19} & \val{ 18}{ 20}{ 16} & \val{1.5}{2.1}{1.0} & \val{ 22}{ 29}{ 18} & \val{2.9}{6.1}{2.1}\\
Leo I & 250 & \vale{1.45}{+0.27}{-0.20} & \val{ 19}{ 21}{ 17} & \val{1.7}{3.1}{1.1} & \val{ 25}{ 27}{ 19} & \val{2.9}{6.3}{2.1}\\
Leo T & 417 & \vale{1.30}{+0.88}{-0.42} & \val{ 16}{ 21}{ 13} & \val{1.2}{2.4}{0.85} & \val{ 19}{ 26}{ 17} & \val{2.4}{6.1}{1.6}\\
\end{tabular}
\caption{The properties of likely DM subhalos of the 18 Milky Way dSph
  galaxies for which $M_{0.3}$ values (column 3) have been published
  \cite{Strigari2008}. $\Vmax$ and $\rVmax$ are the maximum circular
  velocity and its radius, $\Vpeak$ and $\rVpeak$ the largest $\Vmax$
  a subhalo ever acquired and its corresponding radius. The first
  number is the median over all \mbox{\textit{Via Lactea II}} subhalos
  matching the dSph's distance and $M_{0.3}$, the numbers in
  parentheses the 16$^{\rm th}$ and 84$^{\rm th}$ percentiles. (See
  text for details.)}
\label{table1}
\end{table*}
\renewcommand{\tabcolsep}{0.1in}

\begin{table*}
\begin{tabular}{lcccccc}
Name & $D$ & $\lum_{\rm tot}$ & $\lum_{0.3}$ & $\Ftot$ & $\Fc$ \\
     & [kpc] & [$10^6 M_\odot^2 \; {\rm pc}^{-3}$] & [$10^6 M_\odot^2 \; {\rm pc}^{-3}$] & [$10^{-5} M_\odot^2 \; {\rm pc}^{-5}$] & [$10^{-5} M_\odot^2 \; {\rm pc}^{-5}$] \\
\hline \hline
Segue 1 & 23 & \val{2.8}{7.2}{0.93} & \val{2.5}{6.1}{0.89} & \val{ 41}{ 110}{ 14} & \val{ 12}{ 34}{5.6} \\
Ursa Major II & 32 & \val{3.5}{7.2}{2.8} & \val{3.1}{6.1}{2.5} & \val{ 28}{ 56}{ 21} & \val{9.5}{ 18}{7.7} \\
Ursa Minor & 66 & \val{6.2}{9.4}{5.1} & \val{4.7}{7.3}{3.1} & \val{ 11}{ 17}{9.3} & \val{5.2}{8.4}{2.5} \\
Draco & 80 & \val{7.0}{9.9}{6.0} & \val{5.6}{8.2}{3.1} & \val{8.8}{ 12}{7.4} & \val{4.3}{6.4}{1.7} \\
Carina & 101 & \val{7.0}{9.4}{4.8} & \val{5.6}{7.3}{3.5} & \val{5.5}{7.3}{3.7} & \val{3.1}{3.8}{1.6} \\
Wilman 1 & 38 & \val{0.88}{2.9}{0.55} & \val{0.85}{2.7}{0.53} & \val{4.9}{ 16}{3.0} & \val{2.6}{6.4}{1.5} \\
Coma Berenices & 44 & \val{1.2}{2.8}{0.78} & \val{1.1}{2.5}{0.70} & \val{4.8}{ 11}{3.2} & \val{2.5}{5.1}{1.6} \\
Sculptor & 80 & \val{2.9}{3.7}{2.3} & \val{2.5}{3.3}{2.0} & \val{3.7}{4.6}{2.8} & \val{2.0}{2.8}{1.6} \\
Ursa Major I & 106 & \val{3.3}{5.4}{2.3} & \val{2.5}{4.5}{1.9} & \val{2.3}{3.8}{1.6} & \val{1.3}{2.4}{0.91} \\
Fornax & 138 & \val{3.5}{4.4}{3.0} & \val{2.9}{3.3}{2.3} & \val{1.4}{1.8}{1.3} & \val{1.00}{1.2}{0.74} \\
Sextans & 86 & \val{1.2}{2.0}{0.77} & \val{1.1}{1.8}{0.69} & \val{1.3}{2.1}{0.83} & \val{0.86}{1.4}{0.55} \\
Leo II & 205 & \val{4.6}{6.5}{3.8} & \val{3.1}{4.7}{2.1} & \val{0.88}{1.2}{0.73} & \val{0.55}{0.85}{0.37} \\
Canes Venatici I & 224 & \val{4.6}{7.9}{3.8} & \val{3.1}{5.0}{2.3} & \val{0.73}{1.3}{0.60} & \val{0.48}{0.79}{0.35} \\
Leo I & 250 & \val{5.2}{7.9}{3.9} & \val{3.2}{5.4}{2.3} & \val{0.66}{1.0}{0.50} & \val{0.41}{0.73}{0.31} \\
Hercules & 138 & \val{1.4}{2.6}{0.94} & \val{1.2}{2.2}{0.80} & \val{0.57}{1.1}{0.39} & \val{0.42}{0.74}{0.28} \\
Canes Venatici II & 151 & \val{1.2}{2.5}{0.79} & \val{1.1}{2.0}{0.68} & \val{0.44}{0.88}{0.27} & \val{0.33}{0.59}{0.21} \\
Leo T & 417 & \val{3.5}{8.2}{2.4} & \val{2.2}{4.1}{1.7} & \val{0.16}{0.38}{0.11} & \val{0.12}{0.24}{0.093} \\
Leo IV & 158 & \val{0.14}{0.43}{0.063} & \val{0.13}{0.39}{0.060} & \val{0.043}{0.14}{0.020} & \val{0.039}{0.12}{0.018} \\
\end{tabular}
\caption{Estimated luminosities and fluxes for the 18 dSph from
  Table~\ref{table1}. $\lum_{\rm tot}$ is the total luminosity and
  $\lum_{\rm 0.3}$ the luminosity from within the central 0.3
  kpc. $\Ftot = \lum_{\rm tot}/4\pi D^2$ is the total
  flux and $\Fc$ the flux from a central region subtending
  $0.15^\circ$ (about the angular resolution of \Fermi\ above 3
  GeV). The first number is the median over all subhalos matching the
  dSph distance and $M_{0.3}$, the numbers in parentheses are the 16$^{\rm
    th}$ and 84$^{\rm th}$ percentiles. The table is ordered by
  decreasing $\Ftot$.}
\label{table2}
\end{table*}

There are several advantages of known dwarf satellite galaxies as DM
annihilation sources: firstly, the kinematics of individual stars
imply mass-to-light ratios of up to several hundred
\cite{Kleyna2005,Munoz2006,Martin2007,Simon2007}, and hence there is
an a priori expectation of high DM densities; secondly, since we know
their location in the sky, it is possible to directly target them with
sensitive atmospheric Cerenkov telescopes (ACT) such as H.E.S.S.,
VERITAS, MAGIC, and STACEE, whose small field of view makes blind
searches impractical; lastly, our approximate knowledge of the
distances to many dwarf satellites would allow a determination of the
absolute annihilation rate, which may lead to a direct constraint on
the annihilation cross section, if the DM particle mass can be
independently measured (from the shape of the spectrum, for example).

Recent observational progress utilizing the Sloan Digital Sky Survey
(SDSS) has more than doubled the number of known dwarf spheroidal
(dSph) satellite galaxies orbiting the Milky Way
\cite{Willman2005a,Willman2005b,Belokurov2006,Zucker2006a,Zucker2006b,Sakamoto2006,Belokurov2007,Irwin2007,Walsh2007},
raising the total from the 9 ``classical'' ones to 23. Many of the
newly discovered satellites are so-called ``ultra-faint'' dSph's, with
luminosities as low as $1,000 L_\odot$ and only tens to hundreds of
spectroscopically confirmed member stars. Simply accounting for the
SDSS sky coverage (about 20\%), the total number of luminous Milky Way
satellites can be estimated to be at least 70. Taking into account the
SDSS detection limits \cite{Koposov2008} and a radial distribution of
DM subhalos motivated by numerical simulations, this estimate can grow
to several hundreds of satellite galaxies in total
\cite{Tollerud2008,Walsh2009}.

In order to assess the strength of the DM annihilation signal from
these dSph's, it is necessary to have an estimate of the total
dynamical mass, or at least $\Vmax$, of the DM halo hosting the
galaxies. Owing to the extreme faintness of these objects and their
lack of a detectable gaseous component \cite{Grcevich2009}, it has
been very difficult to obtain kinematic information that allows for
such measurements. Progress has been made through spectroscopic
observations of individual member stars, whose line-of-sight velocity
dispersions have confirmed that these objects are in fact strongly DM
dominated \cite{Kleyna2005,Munoz2006,Martin2007,Simon2007}. Such data
best constrain the enclosed dynamical mass within the stellar extent,
which on average is about 0.3 kpc for current data sets. A recent
analysis has determined $M_{0.3} \equiv M(<0.3\;{\rm kpc})$ for 18 of
the Milky Way dSph's, and found that, surprisingly, they all have $M_{0.3}
\approx 10^7 \Msun$ to within a factor of two \cite{Strigari2008}.

State-of-the-art cosmological numerical simulations of the formation
of the DM halo of a Milky Way scale galaxy, such as those of the Via
Lactea Project \cite{Diemand2007a,Diemand2008} and the Aquarius
Project \cite{Springel2008}, have now reached an adequate mass and
force resolution to directly determine $M_{0.3}$ in their simulated
subhalos. This makes it possible to infer the most likely values of
$(\Vmax, \rVmax)$ for a Milky Way dSph of a given $M_{0.3}$ and
Galacto-centric distance $D$, by identifying all simulated subhalos
with comparable $M_{0.3}$ and $D$ and averaging over their $(\Vmax,
\rVmax)$. This analysis was performed for the 9 ``classical'' dwarfs
using the \mbox{\textit{Via Lactea I}} simulation \cite{Madau2008},
and we extend it here to all 18 dwarfs published in
\cite{Strigari2008} and with the more recent and higher resolution
\mbox{\textit{Via Lactea II}} (VL2) simulation.

We randomly generated 100 observer locations at 8 kpc from the VL2
host halo center, and selected, for each Milky Way dSph in
\cite{Strigari2008} separately, all simulated subhalos with distances
within 40\% and numerically determined $M_{0.3}$ within the published
$1-\sigma$ error bars. We then determined the median value and the
16$^{\rm th}$ and 84$^{\rm th}$ percentiles of $(\Vmax, \rVmax)$ and
$(\Vpeak, \rVpeak)$ for each dSph. These values are given in
Table~\ref{table1}. The median values of $\Vmax$ range from 5.0 km
s$^{-1}$ (Leo IV) to 19 km s$^{-1}$ (Draco, Leo I), and of $\Vpeak$
from 6.7 km s$^{-1}$ (Leo IV) to 30 km s$^{-1}$ (Ursa Minor,
Carina). Note that, as expected, dSph's closer to the Galactic Center
typically show a larger reduction from $\Vpeak$ to $\Vmax$, sometimes
by more than a factor of 2.

In the same fashion, we then determine the most likely annihilation
luminosities for the 18 dSph's by using Eq.~(\ref{eq:Lhalo}) for an
NFW profile to calculate the total luminosity $\lum_{\rm tot}$ for every
simulated subhalo. Additionally we also determine $\lum_{0.3}$, the
luminosity within 0.3 kpc from the center, motivated by the fact that
we only have dynamical evidence for a DM dominated potential out to
this radius. Lastly we also consider two measures of the brightness of
each halo: $\Ftot = \lum_{\rm tot}/4\pi D^2$, the total expected
flux from the dSph, and $\Fc$, the flux from a central region
subtending $0.15^\circ$, which is comparable to the angular resolution
of \Fermi\ above 3 GeV. $\Fc$ thus corresponds to the brightest
``pixel'' in a \Fermi\ $\gamma$-ray image of a subhalo. These numbers
are given in Table~\ref{table2}.

\subsection{Current observational constraints}

Several ACT have performed observations of a handful of dSph's.
\begin{itemize}
\item The H.E.S.S. array (consisting of four 107 m$^2$ telescopes with
  a 5$^\circ$ field of view and an energy threshold of $160$ GeV
  \cite{Bernloehr2003}) has obtained an 11 hour exposure of the
  Sagittarius dwarf galaxy. No $\gamma$-ray signal was detected,
  resulting in a flux limit of $3.6 \times 10^{-12}$ cm$^{-2}$
  s$^{-1}$ (95\% confidence) at $E>250$ GeV, and a corresponding limit
  on the cross section of $\sigv \lesssim 10^{-23} \; {\rm cm}^3 \;
  {\rm s}^{-1}$ for an NFW profile and $\sigv \lesssim 2 \times
  10^{-25} \; {\rm cm}^3 \; {\rm s}^{-1}$ for a cored profile (for a
  $\mDM = 100 \; {\rm GeV} - 1 \; {\rm TeV}$ neutralino)
  \cite{Aharonian2008}. Note that the Sagittarius dwarf is undergoing
  a strong tidal interaction with the Milky Way galaxy
  \cite{Martinez-Delgado2004}, and no confident determination of
  $M_{0.3}$ has been possible.
\item The VERITAS array (consisting of four 144 m$^2$ telescopes with
  a 3.5$^\circ$ field of view and an energy threshold of $100$ GeV
  \cite{Holder2008}) has conducted a 15 hours observation of Willman 1
  and 20 hour observations each of Draco and Ursa Minor
  \cite{Hui2008}. No $\gamma$-ray signal was detected at a flux limit
  of $\sim 1\%$ of the flux from the Crab Nebula, corresponding to a
  limit of $2.4 \times 10^{-12}$ cm$^{-2}$ s$^{-1}$ (95\% confidence)
  at $E>200$ GeV \cite{Essig2009}.
\item Additionally the MAGIC \cite{Albert2008, Aliu2009} and STACEE
  \cite{Driscoll2008} telescopes have reported observations of \mbox{Willman 1}
  and Draco, resulting in comparable or slightly higher flux limits.
\end{itemize}
To convert the values of $\Fc$ in Table~\ref{table2} into physical
fluxes that can directly be compared to these observational limits, it
would be necessary to obtain values of the particle physics term of
Eq.~(\ref{eq:Lconversion}) by performing a scan of the DM model
parameter space. This is beyond the scope of this work, but a similar
analysis has been performed by others
\cite{Strigari2008b,Bovy2009,Martinez2009,Pieri2009,Essig2009}. Current
ACT observations of dSph's are beginning to directly constrain DM
models, and future longer exposure time observations of additional
dSph's (in particular Segue 1 and Ursa Major II) with a lower
threshold energy hold great potential. We also eagerly await the first
\Fermi\ data on fluxes from the known dSph galaxies.

\begin{figure*}
\includegraphics[width=\textwidth]{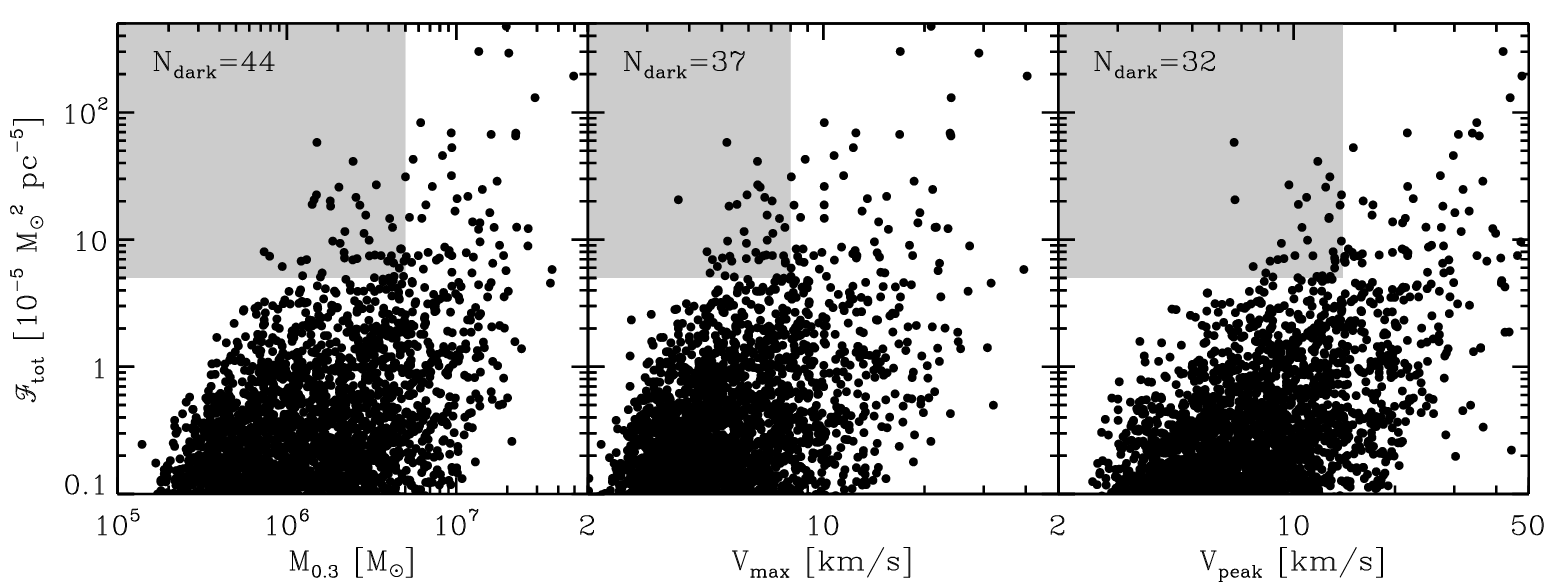}
\caption{The annihilation flux $\Ftot$ from subhalos in the
  \textit{Via Lactea II} simulation versus their $M_{0.3}$, $\Vmax$,
  and $\Vpeak$. The gray shaded areas indicate regions containing
  subhalos with $\Ftot$ as least as high as the fifth-brightest Milky
  Way dSph galaxy (Carina), but with $M_{0.3}, \Vmax, \Vpeak$ below
  that of the known dSph's, i.e. probable dark clumps. Only one of the
  100 random observer locations used in the analysis is shown here.}
\label{fig:Ftot}
\end{figure*}

\section{Dark Clumps}

An annihilation signal from dark clumps not associated with any known
luminous stellar counterpart would provide evidence for one of the
fundamental implications of the CDM paradigm of structure formation:
abundant Galactic substructure. Barring a serendipitous discovery with
an ACT, the discovery of such a source will have to rely on all-sky
surveys, such as provided by \Fermi. Of course even a weak and
tentative identification of a dark clump with \Fermi\ could be
followed up with an ACT.

Unlike for known dSph galaxies, for which we at least have some
astronomical observations to guide us, we must rely entirely upon
numerical simulations to quantify the prospects of detecting the
annihilation signal from dark clumps. Recent significant progress
\cite{Governato2007} notwithstanding, it is at present not yet possible to
perform realistic cosmological hydrodynamic galaxy-formation
simulations, which include, in addition to the DM dynamics, all the
relevant baryonic physics of gas cooling, star formation, supernova
and AGN feedback, etc. that may have a significant impact on the DM
distribution at the centers of massive halos. Instead we make use of
the extremely high resolution, purely collisionless DM-only
\textit{Via Lactea II} (VL2) simulation \cite{Diemand2008}, which
provides an exquisite view of the clumpiness of the Galactic DM
distribution, but at the expense of not capturing all the relevant
physics at the baryon-dominated Galactic center. For the abundance,
distribution, and internal properties of the DM subhalos that are the
focus of this work, the neglect of baryonic physics is less of a
problem, since they are too small to allow for much gas cooling and
significant baryonic effects (this is supported by the high
mass-to-light ratios observed in the Milky Way dSph's), although tidal
interactions with the Galactic stellar and gaseous disk might
significantly affect the population of nearby subhalos.

With a particle mass of $4,100 \Msun$ and a force softening of 40 pc,
the VL2 simulation resolves over 50,000 subhalos today within the
host's $r_{\rm 200}=402$ kpc (the radius enclosing an average density
200 times the mean matter value). Above $\sim 200$ particles per halo,
the differential subhalo mass function is well-fit by a single power
law, $dN/dM \sim M^{-1.9}$, and the cumulative $\Vmax$ function is
$N(>\Vmax) \sim \Vmax^{-3}$ \cite{Diemand2008}. The radial
distribution of subhalos is ``anti-biased'' with respect to the host
halo's density profile, meaning that the mass distribution becomes
less clumpy as one approaches the host's center
\cite{Kuhlen2007,Diemand2008}. Similar results have been obtained by
the Aquarius group \cite{Springel2008, Navarro2008}. Typical subhalo
concentrations, defined as $\Delta_V = \langle \rho(<\rVmax) \rangle /
\rho_{\rm crit}$, grow towards the center, owing to a combination of
earlier formation times \cite{Diemand2005,Moore2006} and stronger
tidal stripping of central subhalos: VL2 subhalos on average have a 60
times higher $\Delta_V$ at 8 kpc than at 400 kpc \cite{Diemand2008}.
Note that this also implies $\sim 7$ times higher annihilation
luminosities for central subhalos, since $\lum \sim \Vmax^4 /\rVmax \sim
\Vmax^3 \sqrt{\Delta_V}$. The counter-acting trends of decreasing
relative abundance of subhalos and increasing annihilation luminosity
towards the center makes it more difficult for (semi-)analytical
methods to accurately assess the role of subhalos in the Galactic
annihilation signal, and motivate future, even higher resolution,
numerical simulations of the formation and evolution of Galactic DM
(sub-)structure. A direct analysis of the VL2 simulations in terms of
the detectability with \Fermi\ of individual subhalos was performed by
\cite{Kuhlen2008}. They found that for reasonable particle physics
parameters a handful of subhalos should be able to outshine the
astrophysical backgrounds and would be detected at more than $5
\sigma$ significance over the lifetime of the \Fermi\ mission.

\begin{figure}
\includegraphics[width=\columnwidth]{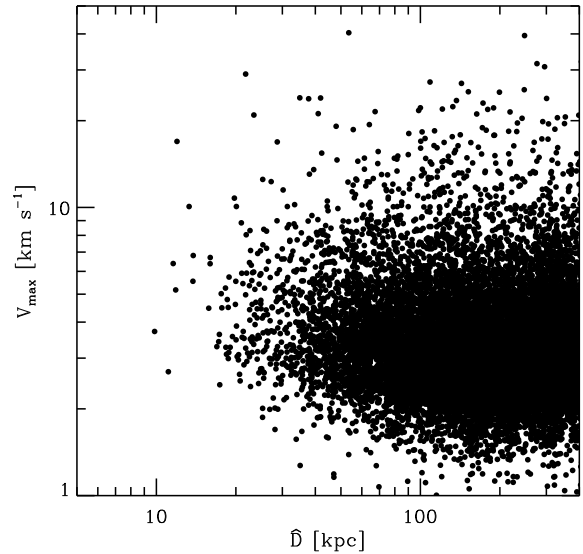}
\caption{VL2 subhalo $\Vmax$ vs. distance to host halo $\hat{D}$.}
\label{fig:Dsub}
\end{figure}

As discussed in the previous section, we have directly calculated the
annihilation luminosities for all VL2 subhalos using
Eq.~(\ref{eq:Lhalo}) and assuming an NFW density profile. The
luminosities would be $\sim 40\%$ \textit{higher} if an Einasto
($\alpha=0.17$) profile had been adopted instead. We then converted
these luminosities to fluxes by dividing by $4\pi D^2$, where the
distances $D$ were determined for 100 randomly chosen observer
locations 8 kpc from the host halo center. The resulting values of
$\Ftot$ are plotted in Figure~\ref{fig:Ftot}, for just one of the 100
observer positions, as a function of the subhalos' $M_{0.3}$, $\Vmax$,
and $\Vpeak$. Although the distributions show quite a bit of scatter,
in all three cases a clear trend is apparent of more massive subhalos
having higher fluxes. This trend could simply be the result of the
higher luminosities of more massive halos, but one might have expected
smaller mass subhalos to be brighter, since their greater abundance
should result in lower typical distances and hence higher fluxes. This
latter effect could be artificially suppressed in the numerical
simulations, if smaller mass subhalos, whose dense centers are not as
well resolved, were more easily tidally disrupted closer to the
Galactic Center, or if the subhalo finding algorithm had trouble
identifying low mass halos in the high background density central
region. In Figure~\ref{fig:Dsub} we plot the subhalos' $\Vmax$ against
their distance to the host halo center $\hat{D}$. There appears to be
a dearth of the lowest $\Vmax$ subhalos ($\Vmax \lesssim 2$ km
s$^{-1}$) at small distances, but at the moment it is not clear
whether this suppression is a real effect or a numerical artifact.
It's also worth noting that such small subhalos might be more
susceptible to disruption by interactions with the Milky Way's stellar
and gaseous disk. At any rate, we can obtain an analytic estimate of
the scaling of the typical subhalo flux with $\Vmax$ by noting that
the luminosity scales as $\lum \sim \Vmax^3 \sqrt{\Delta_V}$ and the
typical distance as $D \sim n^{-1/3} \sim \Vmax^{4/3}$ (since
$dn/d\Vmax \sim \Vmax^{-4}$). The typical flux should thus scale as
$\F \sim \lum/D^2 \sim \Vmax^{1/3} \sqrt{\Delta_V}$, and would be higher
for more massive subhalos at a fixed $\Delta_V$. Actually lower
$\Vmax$ subhalos might be expected to have higher $\Delta_V$ due to
their earlier formation times, but it remains to be seen to what
degree this expectation is borne out in numerical simulations.

\begin{figure}
\includegraphics[width=\columnwidth]{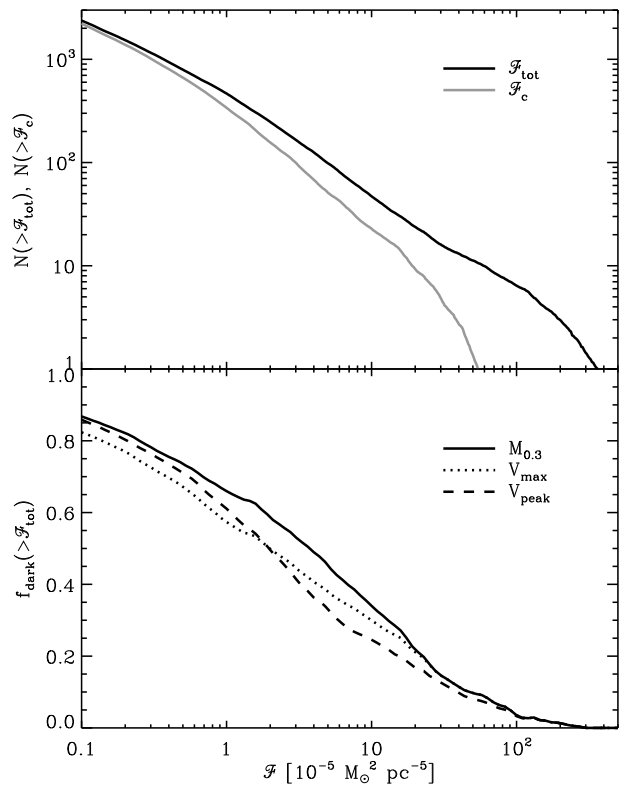}
\caption{\textit{Top:} The cumulative number of subhalos with flux
  exceeding $\Ftot$, $\Fc$. \textit{Bottom:} The fraction of dark
  clumps, i.e. subhalos likely not hosting any stars and defined by
  \mbox{$M_{0.3} < 5\times 10^6 \Msun$}, \mbox{$\Vmax < 8$ km
    s$^{-1}$}, or \mbox{$\Vpeak < 14$ km s$^{-1}$}, as a function of
  limiting flux $\Ftot$. These distributions are averages over 100
  randomly chosen observer locations 8 kpc from the host halo center.}
\label{fig:cumdist+fdark}
\end{figure}

The points in Figure~\ref{fig:Ftot} can be directly compared with the
values for the known Milky Way dSph's in Tables~\ref{table1} and
~\ref{table2}: it appears that there are many DM subhalos at least as
bright as the known Milky Way dSph's. This impression is confirmed by
the top panel of Figure~\ref{fig:cumdist+fdark}, in which we show the
cumulative number of subhalos with fluxes greater than $\Ftot$ and
$\Fc$. These distributions were obtained by averaging over 100
randomly chosen observer locations 8 kpc from the host halo center.
The mean number of DM subhalos with $\Ftot$ greater than that of
(Carina, Draco, Ursa Minor, Ursa Major, Segue 1) is (90, 54, 43, 17,
13), and the corresponding numbers for $\Fc$ are (96, 62, 49, 24, 19).
This demonstrates that if a DM annihilation signal from any of the
known Milky Way dSph's is detected, then many more DM subhalos should
be visible. The plot also implies that Segue 1, the dSph with the
highest $\Ftot$ and $\Fc$ of the currently known sample, is unlikely
to be the brightest DM subhalo in the sky. Of course some of these
additional bright sources could very well have stellar counterparts
that have simply been missed so far, due to the limited sky coverage
of current surveys or insufficiently deep exposures. To assess what
fraction of high flux sources are likely to be genuinely dark clumps
without any stars, we split the sample by a limiting value of
\mbox{$M_{0.3}=5 \times 10^6 \Msun$}, \mbox{$\Vmax=8$ km s$^{-1}$},
and \mbox{$\Vmax=14$ km s$^{-1}$}. We assume that DM subhalos below
these limits are too small to have been able to form any stars, and
hence are truly dark clumps. Of the known dSph's listed in
Table~\ref{table1} only \mbox{Leo IV} falls below these limits. In the
bottom panel of Figure~\ref{fig:cumdist+fdark} we plot $f_{\rm
  dark}(>\Ftot)$, the fraction of subhalos without stars, as a
function of the limiting annihilation flux $\Ftot$. $f_{\rm dark}$
falls monotonically with $\Ftot$, which makes sense given that higher
flux sources are typically more massive and hence more likely to host
stars. Between 30 and 40\% of all DM subhalos brighter than Carina are
expected to be dark clumps. This fraction drops to 10\% for subhalos
brighter than Segue 1.

\subsection{Boost Factor?}

The analysis presented here so far has been limited to known dSph
galaxies and clumps resolved in the VL2 simulation, whose resolution
limit is set by the available computational resources, and has nothing
to do with fundamental physics. Indeed, the CDM expectation is that
the clumpiness should continue all the way down to the cut-off in the
matter power spectrum, set by collisional damping and free streaming
in the early universe \citep{Green2005, Loeb2005}. For typical WIMP
DM, this cut-off occurs at masses of $m_0 = 10^{-12}$ to $10^{-4}
\Msun$ \citep{Profumo2006,Bringmann2009}, some 10 to 20 orders of
magnitude below VL2's mass resolution. Since the annihilation rate
goes as $\rho^2$ and $\langle \rho^2 \rangle > \langle \rho
\rangle^2$, this sub-resolution clumpiness will lead to an enhancement
of the total luminosity compared to the smooth mass distribution in
the simulation.

The magnitude of this so-called substructure boost factor depends
sensitively on the properties of subhalos below the simulation's
resolution limit, in particular on the behavior of the
concentration-mass relation. A simple power law extrapolation of the
contribution of simulated subhalos to the total luminosity of the host
halo leads to boosts on order of a a few hundred
\cite{Springel2008b}. More sophisticated (semi-)analytical models,
accounting for different possible continuations of the
concentration-mass relation to lower masses, typically find smaller
boosts of around a few tens
\cite{Strigari2007,Pieri2008,Kuhlen2008,Martinez2009}.

More importantly, this boost refers to the enhancement of the
\textit{total} annihilation luminosity of a subhalo, but this is not
likely the quantity most relevant for detection. At the distances
where subhalos might be detectable as individual sources, their
projected size exceeds the angular resolution of today's detectors.
The surface brightness profile from annihilations in the smooth DM
component would be strongly peaked towards the center (yet probably
still resolved by \Fermi\ \cite{Kuhlen2008}), owing to the $\rho(r)^2$
dependence of the annihilation rate. The luminosity contribution from
a subhalo's sub-substructure population (i.e. its boost), however, is
much less centrally concentrated: at best it follows the subhalo's
mass density profile $\rho(r)$, although it might very well even be
anti-biased. This implies that substructure would preferentially boost
the outer regions of a subhalo, where the surface brightness typically
remains below the level of astrophysical backgrounds and hence doesn't
contribute much to the detection significance. In other words, the
boost factor might apply to $\Ftot$, but much less (or not at all) to
$\Fc$; yet it is $\Fc$ that is likely to determine whether a given
subhalo can be detected with \Fermi\ or an ACT. It thus seems unlikely
that the detectability of Galactic subhalos would be significantly
enhanced by their own substructure\footnote{This is in contrast to
  many previous claims in the literature, including some by the
  present author \cite[e.g.][]{Kuhlen2008}. A re-analysis of that work
  (in progress) with an improved treatment of the angular dependence
  of the substructure boost, indeed finds that the boost only weakly
  increases the number of detectable subhalos.}. On the other hand, a
substructure boost could be very important for diffuse DM annihilation
signals, either from extragalactic sources, where the boost would
simply increase the overall amplitude \cite{Ullio2002}, or from
Galactic DM, where the boost could affect the amplitude and angular
profile of the signal, as well as the power spectrum of its
anisotropies \cite{Pieri2008, Kuhlen2008, Siegal-Gaskins2008,
  Springel2008b, Fornasa2009, Ando2009}.

\section{Summary and Conclusions}

In this work we have reviewed the DM annihilation signal from Galactic
subhalos. After going over the basics of the annihilation process with
a focus on the resulting $\gamma$-ray output, we summarized the
properties of DM subhalos relevant for estimating their annihilation
luminosity. In the remainder of the paper we used the \textit{Via
  Lactea II} simulation to assess the strength of the annihilation
flux from both known Galactic dSph galaxies as well as from dark
clumps not hosting any stars. By matching the distances $D$ and
central masses $M_{0.3}$ of simulated subhalos to the corresponding
published values of 18 known dSph's, we were able to infer most
probable values, and the 1-$\sigma$ scatter around them, for $\Vmax$
and $\rVmax$, and hence for the annihilation luminosity $\lum$ and flux
$\F$ of all dwarfs. According to this analysis, the recently
discovered dSph Segue 1 should be the brightest of the known dSph's,
followed by Ursa Major II, Ursa Minor, Draco, and Carina. Further, we
showed that if any of the known Galactic dSph's are bright enough to
be detected, then at least 10 times more subhalos should appear as
visible sources. Some of these would be as-of-yet undiscovered
luminous dwarf galaxies, but a significant fraction should correspond
to dark clumps not hosting any stars. The fraction of dark clump
sources is 10\% for subhalos at least as bright as Segue 1 and grows
to 40\% for subhalos brighter than Carina. Lastly, we briefly
considered the role that a substructure boost factor should play in
the detectability of individual Galactic dSph's and other DM
subhalos. We argued that any boost is unlikely to strongly increase
their prospects for detection, since its shallower angular dependence
would preferentially boost the outer regions of subhalos, which
typically don't contribute much to the detection significance.

Several caveats to these findings are in order. Probably the most
important of these is that our simulation completely neglects the
effects of baryons. Gas cooling, star formation, and the associated
feedback processes are unlikely to strongly affect most subhalos,
owing to their low masses. However, tidal interactions with the
baryonic components of the Milky Way galaxy might do so. The
Sagittarius dSph, for example, is thought to be in the process of
complete disruption from tidal interactions with the Milky Way. A
second caveat is that our analysis is based on only one, albeit very
high resolution, numerical simulation, and so we cannot assess the
importance of cosmic variance, or the dependence on cosmological
parameters such as $\sigma_8$ and $n_s$. Other work has found
considerable halo-to-halo scatter
\cite{Reed2005,Springel2008,Ishiyama2009}, with a factor of $\sim 2$
variance in the total subhalo abundance, for example.

These caveats motivate further study and future, higher resolution
numerical simulations, including the effects of baryonic physics. The
characterization of the Galactic DM annihilation signal is of crucial
importance in guiding observational efforts to shed light on the
nature of DM. We are hopeful that in the next few years the promise of
a DM annihilation signal will come to fruition, and will help us to
unravel this puzzle.

\section{Acknowledgments}
Support for this work was provided by the William L. Loughlin
Fellowship at the Institute for Advanced Study. I would like to thank
my collaborators from the Via Lactea Project for their expertise and
invaluable contributions.

\bibliography{dgc}

\end{document}